\documentclass[useAMS,usenatbib,a4paper,referee]{mn2e}
\usepackage{graphicx}
\usepackage{amsmath}
\usepackage{txfonts}
\usepackage{mathrsfs}

\def\aap{A\&A}
\def\apjl{ApJ}

\def\apjs{ApJS}

\newcommand{\be}{\begin{equation}}
\newcommand{\ee}{\end{equation}}
\newcommand{\bea}{\begin{eqnarray}}
\newcommand{\eea}{\end{eqnarray}}

\title[The $R_{\rm h}=ct$ Universe]{On Recent Claims Concerning the $R_{\rm h}=ct$ Universe}
\author[Fulvio Melia]{Fulvio Melia\thanks{John Woodruff Simpson 
Fellow. E-mail: fmelia@email.arizona.edu}\\
\null Department of Physics, The Applied Math Program, and Department of Astronomy, 
The University of Arizona, AZ 85721, USA}

\begin{document}

\date{}

\pagerange{\pageref{firstpage}--\pageref{lastpage}} \pubyear{2013}

\maketitle

\label{firstpage}

\begin{abstract}
The $R_{\rm h}=ct$ Universe is a Friedmann-Robertson-Walker (FRW) cosmology which,
like $\Lambda$CDM, assumes the presence of dark energy in addition to (baryonic
and non-luminous) matter and radiation. Unlike $\Lambda$CDM, however, it is also
constrained by the equation of state (EOS) $p=-\rho/3$, in terms of the total pressure
$p$ and energy density $\rho$. One-on-one comparative tests between $R_{\rm h}=ct$ 
and $\Lambda$CDM have been carried out using over 14 different cosmological 
measurements and observations. In every case, the data have favoured $R_{\rm h}=ct$ 
over the standard model, with model selection tools yielding a likelihood $\sim$$90-
95\%$ that the former is correct, versus only $\sim$$5-10\%$ for the latter. In other 
words, the standard model without the EOS $p=-\rho/3$ does not appear to be the optimal 
description of nature. Yet in spite of these successes---or perhaps because of 
them---several concerns have been published recently regarding the fundamental 
basis of the theory itself. The latest paper on this subject even claims---quite 
remarkably---that $R_{\rm h}=ct$ is a vacuum solution, though quite evidently 
$\rho\not=0$. Here, we address these concerns and demonstrate that all criticisms 
leveled {\it thus far} against $R_{\rm h}=ct$, including the supposed vacuum
condition, are unwarranted. They all appear to be based on incorrect assumptions 
or basic theoretical errors. Nevertheless, continued scrutiny such as this will
be critical to establishing $R_{\rm h}=ct$ as the correct description of nature.
\end{abstract}

\begin{keywords}
{cosmological parameters, cosmology: observations,
cosmology: theory, gravitation}
\end{keywords}

\section{Introduction}
One of the most basic FRW models, $\Lambda$CDM, assumes that the energy density
of the Universe $\rho$ contains matter $\rho_{\rm m}$ and radiation $\rho_{\rm r}$, 
which we see directly, and an as yet poorly understand `dark' energy 
$\rho_{\rm de}$, whose presence is required by a broad range of data including, 
and especially, Type Ia SNe (Riess et al. 1998; Perlmutter et al. 1999). In the
concordance version of $\Lambda$CDM, dark energy is a cosmological constant 
$\Lambda$ with an equation of state (EOS) $w_{\rm de}\equiv w_\Lambda\equiv p_{\rm de}/
\rho_{\rm de}=-1$. For the other two constituents, one simply uses the prescription 
$p_{\rm r}=\rho_{\rm r}/3$ and $p_{\rm m}\approx 0$, consistent with a fully 
relativistic fluid (radiation) on the one hand, and a non-relativistic fluid 
(matter) on the other.

As the measurements continue to improve, however, the EOS $p=w\rho$, where
$w=(\rho_{\rm r}/3-\rho_\Lambda)/\rho$, appears to be creating some tension 
between theory and several observations. The concordance model does 
quite well explaining many of the data, but appears to be inadequate 
to explain all of the nuances seen in cosmic evolution and the growth of 
structure. For example, $\Lambda$CDM cannot account for the general uniformity 
of the CMB across the sky without invoking an early period of inflated expansion 
(Guth 1981; Linde 1982), yet the latest observations with {\it Planck} (Ade et 
al. 2013) suggest that the inflationary model may be in trouble at a fundamental 
level (Ijjas et al. 2013, 2014; Guth et al. 2013). And insofar as the CMB fluctuations
measured with both WMAP (Bennett et al. 2003) and {\it Planck} are concerned,
there appears to be some inconsistency between the predicted and measured
angular correlation function (Copi et al. 2009, 2013; Melia 2014a; Bennett et al.
2013). There is also an emerging conflict between the observed matter distribution 
function, which is apparently scale-free, and that expected in $\Lambda$CDM, 
which has a different form on different spatial scales. The fine tuning required 
to resolve this difference led Watson et al. (2011) to characterize the matter 
distribution function as a `cosmic coincidence.' It also appears that
the predicted redshift-age relation in $\Lambda$CDM's may not be consistent 
with the growth of quasars at high redshift (Melia 2013a), nor the emgergence
of galaxies at high redshift (Melia 2014b).

There is therefore considerable interest in refining the basic $\Lambda$CDM
model, or perhaps eventually replacing it if necessary, to improve the
comparison between theory and observations. Over the past several years,
we have been developing another FRW cosmology, known as the $R_{\rm h}=ct$
Universe, that has much in common with $\Lambda$CDM, but includes an additional
ingredient motivated by several theoretical and observational arguments
(Melia 2007; Melia \& Abdelqadr 2009; Melia \& Shevchuk 2012).
Like $\Lambda$CDM, it also adopts the equation of state $p=w\rho$, with $p=p_{\rm m}+
p_{\rm r}+p_{\rm de}$ and $\rho=\rho_{\rm m}+\rho_{\rm r}+\rho_{\rm de}$, but
goes one step further by specifying that $w=(\rho_{\rm r}/3+
w_{\rm de}\rho_{\rm de})/\rho=-1/3$ at all times. Some observational support for
this constraint is provided by the fact that an optimization of the parameters
in $\Lambda$CDM yields a value of $w$ averaged over a Hubble time equal to
$-1/3$ within the measurement errors. That is, though $w=(\rho_{\rm r}/3-\rho_\Lambda)/\rho$
in $\Lambda$CDM cannot be equal to $-1/3$ from one moment to the next, its value averaged
over the age of the Universe\footnote{It
is not difficult to demonstrate this result. One simply assumes the WMAP
values for the parameters in $\Lambda$CDM and calculates $w(t)$ as a function
of cosmic time from the various contributions to $p$ and $\rho$ due to radiation,
matter, and a cosmological constant. Then averaging $w(t)$ over a Hubble
time, one finds that $\langle w\rangle\approx -0.31$. See the introductory
discussion in Melia (2007) and Melia \& Abdelqader (2009) and, especially,
the more complete description in Melia (2009), particularly figure~1 in this
paper.} is equal to what it would have been in $R_{\rm h}=ct$.

But there are good reasons to believe that $w$ must in fact always be equal 
to $-1/3$ when one uses the FRW metric to describe the cosmic spacetime. This 
metric is founded on the Cosmological principle and Weyl's postulate, which
together posit that the Universe is homogeneous and isotropic (at least on large,
i.e., $>100$ Mpc, spatial scales), and that this high degree of symmetry must be 
maintained from one time slice to the next. Weyl's postulate requires that every
proper distance $R$ in this spacetime be the product of a universal function
of time $a(t)$ (the expansion factor) and a comoving distance $r$. As shown in 
Melia (2007) and Melia \& Shevchuk (2012), the Misner-Sharp mass, given in
terms of $\rho$ and proper volume $4\pi R^3/3$ (Misner \& Sharp 1964), 
defines a gravitational radius $R_{\rm h}=ct$ for the Universe coincident with 
the better known Hubble radius $\equiv c/H$, where $H\equiv \dot{R}/R$ is the 
Hubble constant. Given its definition, $R_{\rm h}=ct$ must itself 
be a proper distance, which trivially leads to the constraint $R_{\rm h}=ct$, 
consistent with an EOS $p=-\rho/3$ (see also Melia \& Abdelqader
2009). As further discussed in Melia (2007) and Melia \& Abdelqader (2009),
the corollary to Birkhoff's theorem, which is of course valid in general
relativity, provides additional justification---and a more pedagogical 
understanding---for defining a spherical {\it proper} volume in which to 
calculate the Misner-Sharp mass. Claims made to the contrary by Bilicki 
\& Seikel (2012) and Mitra (2014) are simply incorrect, and stem from
these authors' misunderstanding of the use of Birkhoff's theorem and
its corollary (see also Weinberg 1972). 

To test whether in fact the EOS $p=-\rho/3$ is be maintained from one
moment to the next, we have carried out an extensive suite of comparative 
tests using $\Lambda$CDM and $R_{\rm h}=ct$, together with a
broad range of observations, from the CMB (Melia 2014a) and high-$z$ quasars
(Melia 2013a, 2014b) in the early Universe, to gamma ray bursts (Wei et al. 2013a) and
cosmic chronometers (Melia \& Maier 2013) at intermediate redshifts and, most recently,
to the relatively nearby Type Ia SNe (Wei et al. 2014a). The total number of tests
is much more extensive than this, and includes the use of time-delay gravitational
lenses (Wei et al. 2014b), the cluster gas-mass fraction (Melia 2013), and the redshift 
dependent star-formation rate (Wei et al. 2014c), among others. In every case, model 
selection tools indicate that the likelihood of $R_{\rm h}=ct$ being correct is 
typically $\sim 90-95\%$ compared with only $\sim 5-10\%$ for $\Lambda$CDM. 
And perhaps the most important distinguishing feature between these two cosmologies 
is that, whereas $\Lambda$CDM cannot survive without inflation, the $R_{\rm h}=ct$ 
Universe does not need it in order to avoid the well-known horizon problem 
(Melia 2013b). 

Yet in spite of the compelling support provided for $R_{\rm h}=ct$ by the
observations, several authors have questioned the validity of this theory.
The earlier claims made by Bilicki \& Seikel (2012) have already been
fully addressed in Melia (2012b), Melia \& Maier (2013), and Wei et al.
(2014a), so we will not revist them here. Similarly, the criticisms
made by van Oirschot et al. (2010) and Lewis (2012) concerning the
definition and use of $R_{\rm h}$ are simply due to their improper 
use of null geodesics in FRW, a full accounting of which was published 
in Bikwa et al. (2012) and Melia (2012a). In this paper, we focus on 
the two most recent claims made concerning the $R_{\rm h}=ct$ 
Universe: (1) that this cosmology is static and merely represents another
vacuum solution (Mitra 2014), and (2) that the equation of state in $R_{\rm h}
=ct$ is inconsistent with $p=-\rho/3$, thus ruining the elegant, high-quality 
fits to the data (Lewis 2013). We will address these two concerns in
\S\S~2 and 3, respectively, and end with some concluding remarks in \S~4.

\section{On Mitra's Claim That $R_{\rm h}=ct$ is a Vacuum Solution}
{Mitra (2014a, and references cited therein) has been trying for 
several years to confirm the validity
and uniqueness of the $R_{\rm h}=ct$ cosmology using the energy complex. 
This is the basis for the claim in his latest paper (Mitra 2014b) that
since $R_{\rm h}=ct$ is (according to him) a vacuum solution, all big 
bang models should be manifestations of the vacuum state as well. 

His argument is based on a presumed demonstration that the 
$R_{\rm h}=ct$ metric is static, for which he then concludes
that $\dot{a}=0$. And since the critical density is proportional
to $\dot{a}$ in the Friedmann equation, he makes the claim that
$R_{\rm h}=ct$ must therefore correspond to a vacuum spacetime.

But his analysis is incorrect for several reasons. First and foremost,
it was proven several decades ago that there are exactly six---{\it and
only six}---special cases of the FRW metric for which a transformation
of coordinates is possible to render the metric coefficients
$g_{\mu\nu}$ $(\mu,\nu=0,1,2,3)$ independent of time $x^0$. These
correspond to solutions of the expansion factor $a(t)$ for which
the spacetime curvature of the FRW metric is constant (Robertson
1929; Florides 1980; Melia 2012c, 2013c). As shown by Florides (1980)
in his landmark paper, these special cases are (1) the Minkowski spacetime,
(which is highly trivial), (2) the Milne Universe (with spatial
curvature constant $k=-1$), (3) de Sitter space, (4) anti-de Sitter space, 
(5) an open Lanczos-like Universe, and (6) the Lanczos Universe itself. 
The $R_{\rm h}=ct$ Universe, with $a(t)\propto t$ and $k=0$, is not
one of them.

The spacetime curvature in $R_{\rm h}=ct$ is {\it not constant} and the
reason $\dot{a}=$ constant in this cosmology is not because $\rho=0$
but, rather, because $\rho+3p=0$---i.e., the `active mass' is zero
(Melia 2014c). Mitra's derivation in \S~3 of his paper is flawed 
because he assumes that the FRW metric can always be written in
`Schwarzschild coordinates.' But this too is incorrect because
the transformed time $T$ (measured from the big bang) is well defined
in only a few special cases, as demonstrated several years ago by Melia
\& Abdelqader (2009). It is not possible to rewrite the FRW metric
solely in terms of R and T in those cases where $R$ can exceed
$R_{\rm h}$, which certainly happens at early times for cosmologies,
such as $\Lambda$CDM and $R_{\rm h}=ct$, with an initial singularity. 
(The de Sitter Universe is an obvious counter-example.)

For these reasons, it is simply wrong for Mitra to claim that the 
$R_{\rm h}=ct$ Universe is merely another manifestation of the vacuum 
solution. 

\section{On Lewis's Claim Concerning the EOS in $R_{\rm h}=ct$}
For reasons that are never made clear, Lewis (2013) assumes
that a `pure' $R_{\rm h}=ct$ Universe is comprised of a single fluid
(dark energy) with no matter, and an equation of state $w=w_{\rm de}=-1/3$.
He then makes the additional assumption that if matter were to be
introduced into such a universe, it ought to be conserved separately
from all the other constituents. Not only are these assumptions
unnecessary, but there is actually no precedent for them either.
In fact, they are incorrect from the outset.

As described above, this is not how the $R_{\rm h}=ct$ Universe is 
set up. As noted earlier, the $R_{\rm h}=ct$ Universe is $\Lambda$CDM 
with the additional constraint $w=-1/3$. This does not mean that 
$w_{\rm de}=-1/3$, nor that $\rho_{\rm m}=\rho_{\rm r}=0$. The models 
considered by Lewis should therefore be more aptly viewed as variants 
of $\Lambda$CDM, and we already know that in order for the standard
model to have any hope of fitting the data, one must have $\Omega_{\rm m}\equiv
\rho_{\rm m}(t_0)/\rho(t_0)\sim 0.27$ and (with analogous definitions)
$\Omega_{\rm de}\equiv\Omega_\Lambda \sim 0.73$, with the spatial flatness
condition $\Omega_{\rm m}+\Omega_{\rm r}+\Omega_\Lambda=1$. It is hardly 
surprising, then, that the unusual models considered by Lewis do not 
fit the data. They are neither $R_{\rm  h}=ct$ nor the concordance
model either.

Second, the assumption of a separately conserved matter field
is not used in FRW cosmologies, and is certainly not valid 
over the age of the Universe in $\Lambda$CDM. There is therefore
no precedent for imposing it on $R_{\rm h}=ct$ either. For example, 
$\Lambda$CDM invokes the idea that matter in the early Universe
was created and annihilated, exchanging its energy density with 
that of the radiation field (and possibly other fields that may 
emerge from extensions of the standard model of particle physics). 
In $\Lambda$CDM, matter may be separately conserved today if 
interations such as these are currently inconsequential. However, 
we don't even know if dark matter is self-interacting, or if it 
decays. So matter could not have been separately conserved in the 
early Universe; it may not even be so conserved today, and may in 
fact never be conserved if its interactions with other energy 
fields continue indefinitely into the future. What we can say for 
sure in the case of $R_{\rm h}=ct$ is that in order for the equation 
of state to be maintained at $w = -1/3$, the various constituents
must adjust their relative densities via particle-particle
interactions. But there is nothing mysterious 
about a situation such as this, in which the internal `chemistry' 
of a system is controlled by external or global physical
constraints. We do the same thing in the standard model when we
force the temperature to obey a fixed functional dependence $T(z)$
on the redshift, and then require all the particle species to find
their equilibrium through the various forces and interactions they
experience with other components. In situations such as this,
it is important to remember that particle numbers are not
conserved, and each particle type is subject to the pressure
of other species, not just its own, so one cannot naively assume
that each component evolves as an independent density. 
For example, it is not correct to assume that prior to recombination, 
when matter and radiation were in local thermodynamic equilibrium, 
the matter energy density scaled as $\rho_{\rm m}\sim a^{-3}$ and 
the radiation as $\rho_{\rm r}\sim a^{-4}$. These only apply
when matter and radiation evolve independently of each other. 

This analogy may appear to be over-reaching; after all, the
spectrum of the CMB is a spectacular Planck function.
But there are already several indicators, some circumstantial,
that the condition $w=-1/3$ is also being maintained as the
Universe expands. We have already alluded to the fact that
$<w>\approx -1/3$ when $w(t)$ is averaged over a Hubble
time. Such an average can emerge only once, in the entire 
history of the Universe---unless $w$ were always equal to
$-1/3$. Otherwise, it would be an extraordinay coincidence 
for us to be living just at this moment, the only instant when 
we can see this happen. This condition is also suggested 
by model-independent measurements
of the Hubble constant $H(z)$ (Melia \& Maier 2013), which are
most consistent with $w=-1/3$ (i.e., $a[t]\propto t$), which
results in $H(z)=H_0(1+z)$. And a more substantial analysis
of the cosmic equation of state yields a strong correlation
between the inferred values of $\Omega_{\rm m}$ and
$w_{\rm de}$ when optimizing the parameters in $\Lambda$CDM
to fit the data (Melia 2014d). This correlation predicts that
$\Omega_{\rm m}\approx 0.27$ when $w_{\rm de}=-1$, while
$w_{\rm de}$ must be closer to -1.1 if $\Omega_{\rm m}\approx 
0.31$.  Interestingly, the first pair of values corresponds
to the WMAP results (Bennett et al. 2003), while the latter 
pair corresponds to the best fit using the Planck measurements 
(Ade et al. 2013). This is still only circumstantial evidence at
best, but it does suggest that the optimization of the parameters
in $\Lambda$CDM is always restricted by the condition $w=-1/3$.

\section{Conclusion}
In spite of the many successes $\Lambda$CDM has enjoyed in accounting
for the cosmic expansion, many today would agree that the ever-improving 
measurements are starting to reveal some possible inconsistencies between 
its predictions and the latest observations. We have highlighted several of
these areas and the need to evolve the standard model in order to address 
these potential problems.

The $R_{\rm h}=ct$ Universe is essentially $\Lambda$CDM with one additional
constraint---the total EOS $p=-\rho/3$. This condition, which is motivated
by several observational and theoretical arguments, appears to solve
many of the conflicts otherwise experienced by the standard model. 
As of today, every one-on-one comparison carried out between these two 
models has statistically favoured the former. It is difficult to argue
against this rate of success. 

Nevertheless, the development of $R_{\rm h}=ct$ as a comprehensive
description of nature is hardly complete, inviting several concerted
efforts at challenging its fundamental basis. Such scrutiny is an
essential component of any serious discussion concerning its viability.
As of today, however, all the criticisms raised thus far appear to have
been based on incorrect assumptions or flawed theoretical arguments.
In this paper, we have discounted the two most recent claims, one
having to do with the presumed vacuous nature of the $R_{\rm h}=ct$
metric, and the second with its possibly inconsistent equation of
state.

\section*{Acknowledgments}
I am grateful to the anonymous referee for a thorough and helpful review.
Part of this work was carried out with the support of the Simpson 
visiting chair at Amherst College, grant 2012T1J0011 from The Chinese 
Academy of Sciences Visiting Professorships for Senior International 
Scientists, and grant GDJ20120491013 from the Chinese State Administration 
of Foreign Experts Affairs at Purple Mountain Observatory in Nanjing, China.


\newpage
\begin{thebibliography}{99}

\bibitem[]{Ade2013} Ade, P.A.R. et al. 2013, A\&A, in press (arXiv:1303.5083)
\bibitem[]{Bennett2003} Bennett, C. L. et al. 2003, ApJ, 583, 1
\bibitem[]{Bennett2013} Bennett, C. L. et al. 2013, ApJS, 208, 20
\bibitem[]{Bikwa2012} Bikwa, O., Melia, F. \& Shevchuk, A.S.H. 2012, MNRAS, 421, 3356
\bibitem[]{Bilicki2012} Bilicki, M. \& Seikel, M. 2012, MNRAS, 425, 1664
\bibitem[]{Birkhoff23} Birkhoff, G. 1923, {\it Relativity and Modern Physics} (Cambridge, Harvard University Press)
\bibitem[]{Copi2009} Copi, C. J., Huterer, D., Schwarz, D. J. \& Starkman, G. D. 2009, MNRAS, 399, 295
\bibitem[]{Copi2013} Copi, C. J., Huterer, D., Schwarz, D. J. \& Starkman, G. D. 2013, MNRAS, in press (arXiv:1310.3831)
\bibitem[]{florides80} Florides, P. S. 1980, GRG, 12, 563
\bibitem[]{Guth1981} Guth, A. H. 1981, PRD, 23, 347
\bibitem[]{Guth2013} Guth, A. H. Kaiser, D. I. \& Nomura, I. 2013, eprint arXiv:1312.7619
\bibitem[]{Ijjas2013} Ijjas, A., Steinhardt, P. J. \& Loeb, A. 2013, PLB, 723, 261
\bibitem[]{Ijjas2014} Ijjas, A., Steinhardt, P. J. \& Loeb, A. 2014, eprint arXiv:1402.6980
\bibitem[]{Lewis2012} Lewis, G. F. \& van Oirschot, P., 2012, MNRAS Let., 423, 26
\bibitem[]{Lewis2013} Lewis, G. F. 2013, MNRAS, 432, 2324
\bibitem[]{Linde1982} Linde, A. 1982, PLB, 108, 389
\bibitem[]{Melia07} Melia, F. 2007, MNRAS, 382, 1917
\bibitem[]{Melia09} Melia, F. 2009, IJPM-D, 18, 1
\bibitem[]{Melia2012a} Melia, F. 2012a, JCAP, 09, 029
\bibitem[]{Melia2012b} Melia, F. 2012b, AJ, 144, 110
\bibitem[]{Melia2012c} Melia, F. 2012c, MNRAS, 422, 1418
\bibitem[]{Melia2013a} Melia, F. 2013a, ApJ, 764, 72
\bibitem[]{Melia2013b} Melia, F. 2013b, A\&A, 553, id A76
\bibitem[]{Melia2013c} Melia, F. 2013c, CQG, 30, 155007
\bibitem[]{Melia2014a} Melia, F. 2014a, A\&A, 561, 80
\bibitem[]{Melia2014b} Melia, F. 2014b, AJ, 147, 120 
\bibitem[]{Melia2014c} Melia, F. 2014c, EPJ-C, submitted
\bibitem[]{Melia2014d} Melia, F. 2014d, PRD, submitted
\bibitem[]{MeliaAbdelqader09} Melia, F. \& Abdelqader, M. 2009, IJMP-D, 18, 1889
\bibitem[]{MeliaMaier13} Melia, F. \& Maier, R. S. 2013, MNRAS, 432, 2669
\bibitem[]{MeliaShevchuk12} Melia, F. \& Shevchuk, A. 2012, MNRAS, 419, 2579
\bibitem[]{Misner64} Misner, C. W. \& Sharp, D. H. 1964, Phys Rev, 136, 571
\bibitem[]{Mitra2014a} Mitra, A. 2014a, New Astronomy, 30, 46
\bibitem[]{Mitra2014b} Mitra, A. 2014b, MNRAS, 442, 382
\bibitem[]{Perlmutter1999} Perlmutter, S. et al. 1999, ApJ, 517, 565
\bibitem[]{Riess1998} Riess, A. G. et al. 1998, AJ, 116, 1009
\bibitem[]{van2010} van Oirschot, P., Kwan, J. \& Lewis, G. F. 2010, MNRAS, 404, 1633
\bibitem[]{Robertson29} Robertson, H. P. 1929, Proceedings of the National Academy of Sciences of 
the United States of America, 15, 822
\bibitem[]{Watson2011} Watson, D. F., Berlind, A. A. \& Zentner, A. R. 2011, ApJ, 738, article id. 22
\bibitem[]{Wei13a} Wei, J.-J., Wu, X.-F. \& Melia, F. 2013a, ApJ, 772, 43
\bibitem[]{Wei14a} Wei, J.-J., Wu, X.-F., Melia, F. \& Maier, R. S. 2014a, AJ, submitted
\bibitem[]{Wei14b} Wei, J.-J., Wu, X.-F. \& Melia, F.  2014b, ApJ, 788, id. 190
\bibitem[]{Wei14c} Wei, J.-J., Wu, X.-F., Melia, F., Wei, D.-M. \& Feng, L.-L. 2014c, MNRAS, 439, 3329
\bibitem[]{Weinberg72} Weinberg, S. 1972, {\it Gravitation and Cosmology: Principles and Applications
of the General Theory of Relativity} (Wiley, New York)

\end{thebibliography}
\end{document}